# Tunneling Breakdown of a Strongly Correlated Insulating State in $VO_2$ Induced by Intense Multi-Terahertz Excitation

B. Mayer,[1] C. Schmidt,[1] A. Grupp,[1] J. Bühler,[1] J. Oelmann,[1] R. E. Marvel,[2] R. F. Haglund Jr.,[2] T. Oka,[3] D. Brida,[1] A. Leitenstorfer,[1,†] and A. Pashkin[1,*]

[1] Department of Physics and Center for Applied Photonics, University of Konstanz, D-78457 Konstanz, Germany.

[2] Department of Physics and Astronomy, Vanderbilt University, Nashville, Tennessee 37235, USA.

[3] Department of Applied Physics, University of Tokyo, Bunkyo, Tokyo 113-8656, Japan.

[†] Corresponding author: aleitens@uni-konstanz.de

We directly trace the near- and mid-infrared transmission change of a $VO_2$ thin film during an ultrafast insulator-to-metal transition triggered by high-field multi-terahertz transients. Non-thermal switching into a metastable metallic state is governed solely by the amplitude of the applied terahertz field. In contrast to resonant excitation below the threshold fluence, no signatures of excitonic self-trapping are observed. Our findings are consistent with the generation of spatially separated charge pairs and a cooperative transition into a delocalized metallic state by THz field-induced tunneling. The tunneling process is a condensed-matter analogue of the Schwinger effect in nonlinear quantum electrodynamics. We find good agreement with the pair production formula by replacing the Compton wavelength with an electronic correlation length of 2.1 Å.

[*] Current address: Helmholtz-Zentrum Dresden-Rossendorf, Institute of Ion Beam Physics and Materials Research, 01328 Dresden, Germany.

# I. INTRODUCTION

Ultrafast electron dynamics in condensed phases is an ideal test bed for studying universal quantum nonequilibrium processes using pump-probe spectroscopy. Recent examples include the observation of the Anderson-Higgs amplitude mode in a BCS superconductor [1,2], where concepts common to condensed-matter and high-energy physics were examined. The Schwinger effect [3,4] also has roots in both fields. In quantum electrodynamics (QED), the Dirac sea is known to be unstable in strong electric fields since electron-positron dipole pairs that exist as virtual quantum fluctuations can gain energy from the field and turn into real excitations. This process, formally equivalent to Zener breakdown in semiconductors [5], has seen renewed attention due to its possible realization in graphene [6,7].

There is much interest in many-body generalizations of the Schwinger effect in systems that undergo metal-insulator transitions due to strong Coulomb interactions. In condensed matter, dielectric breakdown in strongly correlated electron systems has been studied theoretically [8-11]. A doublon (doubly occupied lattice site)-holon (empty site) pair is created when the field exceeds the threshold [11]

$$E_\text{th} = \frac{\Delta}{2e\xi} \qquad (1)$$

called the Schwinger limit, where $\Delta$ is the charge gap, $e$ the unit charge and $\xi$ the correlation length [12,13]. The latter parameter corresponds to the typical spatial extent of doublon-holon pairs and plays a role similar to the coherence length in superconductors. In fact, one can use Eq. (1) to estimate the correlation length by dielectric breakdown experiments [14]. In high-energy physics, the Schwinger effect is relevant to quark-gluon plasma formation in heavy ion collisions [15], quark-antiquark pair production [16,17] and subsequent deconfinement, which corresponds to an insulator-to-metal transition (IMT) studied using string theory techniques [18]. This progress motivates direct experiments on the many-body Schwinger effect. Studies carried out in a strongly correlated condensed-matter system thus enrich our general understanding of nonequilibrium quantum physics.

Vanadium dioxide ($VO_2$) is a prime example of a strongly correlated material with an insulator-to-metal transition (IMT) at $T_c = 340$ K accompanied by a structural transformation from a monoclinic into a rutile crystal structure by dimerization of the vanadium sublattice [19].

Although quasi-static electric fields can induce an IMT in $VO_2$ in certain cases, the Schwinger effect is dominated by other physical mechanisms. Studies of the voltage-controlled IMT in thin films show that even high DC fields require Joule heating to complete the transition [20]. A recent report of ionic-liquid gating of the IMT by electrostatic fields [21] appears to result from the creation of oxygen vacancies [22] as previously observed in nanoparticles [23]. Thus, a transient electric bias provided by ultrashort terahertz (THz) pulses paves a new way towards the observation of true many-body tunneling in correlated systems. A first demonstration of field-induced metallization of $VO_2$ inside a metamaterial structure was reported recently using picosecond THz waveforms with local fields exceeding 1 MV/cm [24]. The observed IMT timescale of 8 ps was interpreted as field-induced enhancement of carrier density (Poole-Frenkel effect) succeeded almost immediately by Joule heating due to electron-lattice coupling, which then stabilizes the metallic phase. However, a truly non-thermal field-driven IMT in $VO_2$ should occur on much faster timescales.

Here we demonstrate efficient induction of the IMT in a $VO_2$ thin film by non-resonant excitation with high-field multi-THz transients with amplitudes as high as 15 MV/cm, on a timescale below 100 fs. Ultrafast switching into a metastable metallic state unequivocally indicates a non-thermal tunneling mechanism. The concentration of the metallic phase in the film depends solely on the peak electric field of the multi-THz pulses and increases rapidly above about 8 MV/cm. Remarkably, in contrast to resonant excitation with near-infrared (NIR) laser pulses, long-lived metallic domains are generated even by THz transients with peak fields well below this value. Our observation is consistent with the generation of non-local doublon-holon pairs via field-induced tunneling that efficiently prevents excitonic self-trapping.

**II. EXPERIMENT**

The studied sample is a 200-nm-thin film of polycrystalline $VO_2$ grown on a diamond substrate by pulsed laser deposition [25]. Fig. 1(a) shows the temperature dependence of the normalized near-infrared transmittance through a 200-nm-thick $VO_2$ film which exhibits a pronounced hysteresis. The IMT occurs at $337 \pm 0.5$ K (64 $^o$C) upon heating and at $330 \pm 0.5$ K (57 $^o$C) on cooling down. The narrow hysteresis width (7 K) attests to the high quality of the thin-film sample. All measurements presented in this paper were acquired at room temperature (294 K). Fig. 1(b) shows an image of the entire $VO_2$ film obtained by merging multiple optical microscope images taken in reflection mode.

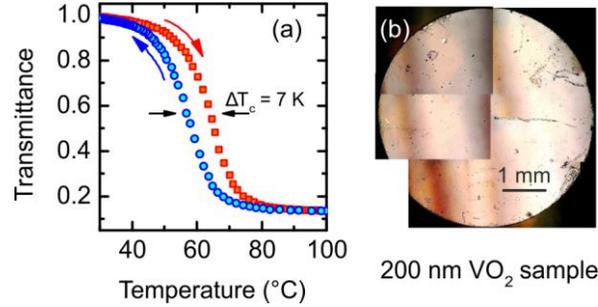

FIG. 1. (a) Temperature dependence of the normalized transmittance through a 200 nm $VO_2$ film upon heating (red) and cooling (blue). (b) Optical microscope image of the 200-nm-thick $VO_2$ layer deposited on a diamond substrate with a diameter of 5 mm.

High-field multi-THz transients at a repetition rate of 1 kHz are generated using difference frequency mixing in a GaSe emitter [26,27]. Real-time field profiles of these phase-stable pulses are detected by electro-optic sampling. Figs. 2(a) and 2(b) show typical waveforms and amplitude spectra of the multi-THz pump pulses, respectively. The pulse duration and spectral bandwidth depend on the thickness of the GaSe emitter [26]. Typical transients used in our experiments comprise only a few electromagnetic cycles oscillating at center frequencies around 20 THz. Owing to short pulse durations and strong focusing by an off-axis parabolic mirror (focal length of 25 mm), we generate peak electric fields up to 15 MV/cm inside the $VO_2$ film. The time-resolved changes in the $VO_2$ are monitored using (i) broadband NIR pulses of 8 fs duration centered at 1.2 μm and (ii) broadband multi-THz transients with 100 fs duration (FWHM) at 25 THz. More details about the experimental setups are given in the Supplementary Material [25].

II. RESULTS AND DISCUSSION

A. THz pump / NIR probe measurements

The differential NIR transmission $\Delta T/T$ induced by an excitation transient with a peak electric field of 6.1 MV/cm is depicted in Fig. 2(c). The differential transmission exhibits a step-like decrease on a timescale of 100 fs, close to the duration of the excitation pulse. The rapid switching highlights the non-thermal character of the THz-driven change in the $VO_2$ dielectric function; the transmission is suppressed for many picoseconds after THz excitation (see also Ref. [25]).

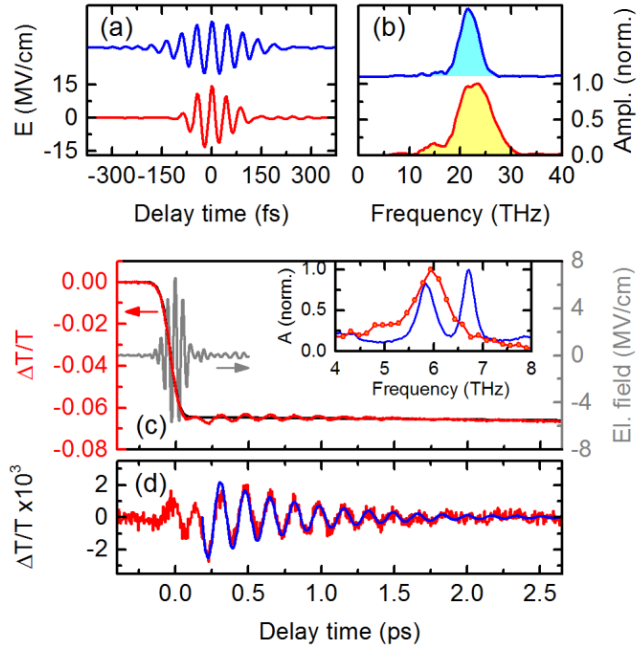

FIG 2. (a) Time-domain profiles of few-cycle multi-THz pump transients generated by 230 μm (red) and 1050 μm (blue) GaSe emitter crystals, respectively. (b) Amplitude spectra of the THz transients from panel (a). Blue graphs are vertically shifted for better visibility. (c) Relative transmission change ΔT/T of the $VO_2$ film (orange) induced by a multi-THz transient (gray) with a peak electric field of 6.1 MV/cm and probed by the NIR pulse. The black solid line represents a fit to the step-like non-oscillating part of the pump-probe signal. Inset: Unpolarized Raman spectrum of the sample (blue) and Fourier transform of the oscillating part of ΔT/T (red circles). (d) Oscillating part of the signal (red) fitted by a damped harmonic oscillation (blue) with a frequency of 5.9 THz.

Another prominent aspect of the THz pump / NIR probe response is the oscillatory feature arising immediately after excitation. When the non-oscillating part of the pump-probe signal shown in Fig. 2(c) is subtracted, the remaining component depicted in Fig. 2(d) fits a single-frequency damped harmonic oscillation. The center frequency of 5.9 THz coincides with the coherent lattice motion in $VO_2$ observed in previous NIR pump / THz probe experiments [28,29].

We emphasize that the symmetry of the monoclinic phase should lead to two distinct tilting modes of the vanadium dimers. Indeed, the spontaneous Raman spectra of our sample shown in the inset of Fig. 2(c) clearly exhibit two sharp modes related to the monoclinic structure of the VO$_2$ film in its ground state. In contrast, Fourier analysis of the coherent lattice oscillation observed after nonlinear excitation reveals only a single mode with significantly faster damping [inset of Fig. 2(c)]. As in the case of NIR excitation above threshold [30], this observation provides evidence for changes of lattice symmetry caused by bond-breaking of the V-V dimers in VO$_2$ induced by the strong THz field.

To elucidate the mechanism of the THz-driven IMT we systematically vary the peak electric field of the pump pulses between 3 and 15 MV/cm. Figs. 3(a) and 3(b) show the time evolution of the differential transmission following excitation by broadband and narrowband multi-THz transients of differing field strengths. In both cases the absolute value |ΔT/T| changes by almost two orders of magnitude while the electric field varies only by a factor of five. As expected for excitation far from resonance, the differential transmission shows a highly nonlinear dependence on the applied THz field, although the ultrafast dynamics is qualitatively similar for broadband and narrowband multi-THz excitations. Clearly, the longer duration of the narrowband transient results in a slower onset time of 190 fs while the smaller bandwidth (4.5 THz vs. 7.7 THz) precludes impulsive excitation of the coherent oscillation at 5.9 THz.

Most importantly, we observe an appreciable difference in signal amplitude depending on the multi-THz pump fluence. Fig. 3(c) and its inset depict the amplitude of the step-like decrease of ΔT/T as a function of peak electric field and excitation fluence, respectively. The shorter, broadband multi-THz transients excite VO$_2$ more efficiently than longer narrowband pulses with the same fluence. However, when the same data are plotted as functions of the peak electric field, the two graphs almost coincide [Fig. 3(c)]. This indicates that the volume fraction undergoing excitation-induced switching is controlled by the electric field of the multi-THz transients and not by the flux of incident THz photons.

The sub-cycle time resolution of our experiment reveals another fundamental difference between NIR and multi-THz pumping. For resonant interband excitation at 800 nm, fast sub-picosecond relaxation towards an insulating state is observed below the threshold fluence for the rutile metallic phase [black graphs in Figs. 3(a)-(b)] [28-31]. This behavior is completely absent for *any* field strength of multi-THz excitation indicating a qualitatively different microscopic mechanism for the non-resonant photoinduced IMT.

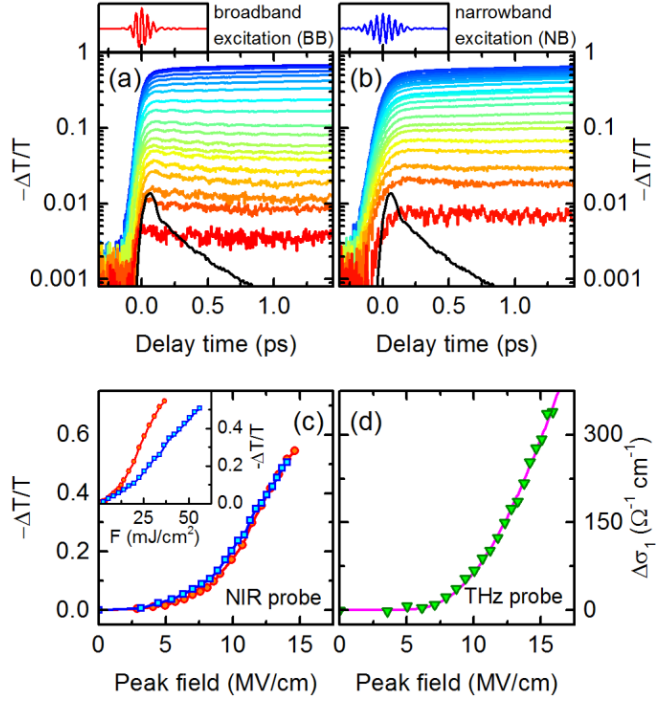

FIG. 3. (a,b) Relative transmission change ΔT/T induced by multi-THz excitation at various peak electric fields and probed with an 8-fs-short NIR pulse. (a) Broadband excitation with peak electric fields ranging from 2.9 MV/cm (red) to 14.6 MV/cm (blue) and (b) narrowband excitation between 3.1 MV/cm (red) and 14.0 MV/cm (blue). The black solid line shows the transient change of multi-THz conductivity in arbitrary units after excitation with a 12 fs pulse at a wavelength of 800 nm and fluence of 3 mJ/cm² taken from Ref. [28]. (c) ΔT/T immediately after excitation as a function of the peak driving field for broadband (red circles) and narrowband (blue squares) pump transients. Inset: same data as a function of incident excitation fluence F. (d) Absolute change in the real part of optical conductivity $\Delta\sigma_1$ at 1 ps pump-probe delay time measured by two-dimensional multi-THz spectroscopy for various driving peak fields (green triangles). The solid line shows a best fit to the data based on Eq. (2).

## B. Excluding the effect of phonon pumping

In order to check the influence of possible resonant excitation of high-frequency optical phonon modes in $VO_2$ we vary the carrier frequency of the multi-THz transients. The absorption coefficient of our $VO_2$ film measured by a Fourier-transform spectrometer [Fig. 4(a)] is dominated by two contributions: (i) infrared-active phonon modes between energies of 34 meV and 77 meV [32]; (ii) interband absorption for photon energies above the band gap of 0.6 eV [33]. This IR absorption spectrum is very similar to that of a bulk $VO_2$ single crystal [33] demonstrating the high quality of the studied thin film. In particular, our sample does not show a pronounced additional absorption feature above the phonon resonances which has been found in thin films used for a previous study on excitation of $VO_2$ with intense mid-infrared pulses [34].

Typical spectra of the multi-THz transients applied for excitation in our experiments are depicted in Fig. 4(b). We tune from strong overlap with the highest oxygen vibrational mode at 18.8 THz all the way to a center frequency of 30 THz which is completely non-resonant. The observed pump-probe response is independent of the central frequency of the multi-THz transients with nominally identical peak electric fields [see Fig. 4(c)]. Broadband excitation centered at 20 THz (red) only results in a faster onset time because the pulse duration is shorter as compared to the narrowband pump pulse centered at 30 THz (blue). But the following dynamics of the induced transmission change is completely identical even for those two limiting cases demonstrating the onset of the metastable metallic phase. The linear absorption due to optical phonons differs by at least one order of magnitude for frequencies of 20 THz and 30 THz, respectively. In contrast, the level of relative transmission change after the excitation is completed shows a variation of less than 20% which is less than the relative error for the determination of the internal peak electric fields at different center frequencies. Thus, we conclude that the resonant infrared excitation of the polar lattice vibration plays a negligible role in the field-induced IMT in $VO_2$.

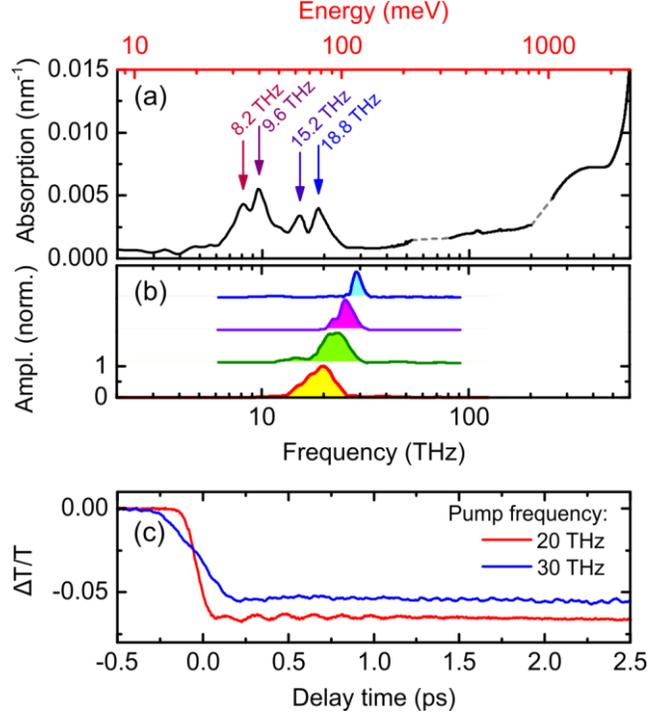

FIG. 4: (a) Absorption coefficient of the 200-nm-thick $VO_2$ film. Arrows mark the frequency of infrared-active phonons. The dashed lines are interpolations in the frequency ranges affected by an absorption in the diamond substrate and a detector response. (b) Typical spectra of the multi-THz excitation transients used in our experiments. The graphs are shifted vertically for clarity. (c) Pump-probe signals for two different central frequencies of the multi-THz excitation pulse with a peak electric field of approximately 7 MV/cm. The red graph corresponds to the broadband spectrum centered at 20 THz which is fully resonant to the highest vibrational mode of the oxygen cage. The blue graph results from narrowband excitation centered at 30 THz which is completely off-resonant with respect to any vibrational modes.

### C. THz pump / THz probe measurements

In order to investigate the origin of the differences between NIR and THz pumping discussed in section II.A., we perform a THz pump / THz probe experiment. Although the NIR probe pulses used in our experiment are sensitive to the IMT in $VO_2$, they cannot unambiguously confirm the metallic character of the excited state: the photon energy of 1 eV mainly traces changes in the density of states well above the insulating energy gap [35]. Therefore, we employ multi-THz transients centered at 25 THz (photon energy of 100 meV) to probe conductivity changes *far below* the gap energy while retaining sub-cycle temporal resolution. The strong suppression of optical transmission at these probe energies clearly marks the closure of the

— Page 9 of 20 —

energy gap and the onset of the Drude-like response of mobile electrons in metallic $VO_2$ [28,32,33]. At a pump-probe delay of $t_D = 1$ ps, we measure the pump-induced changes in the complex transmission coefficient to evaluate directly the changes in the complex optical conductivity $\Delta\sigma(\omega)$ around 25 THz [25]. Fig. 3(d) shows the spectrally averaged real part of $\Delta\sigma(\omega)$ as a function of the peak excitation field. There is a clear excitation threshold: $\Delta\sigma_1$ vanishes for excitation peak fields below approximately 8 MV/cm and starts to grow rapidly for higher applied fields. This behavior is more pronounced than the response in the NIR range [Fig. 3(c)], since THz probe is much more sensitive to metallization of $VO_2$ [28,36]. The existence of an excitation threshold provides decisive evidence that the IMT in $VO_2$ is induced by high-field THz transients. At the highest internal field of 15 MV/cm the real part $\Delta\sigma_1$ exceeds 300 $\Omega^{-1}cm^{-1}$, almost double the conductivity change reported under NIR pumping [28]. This fact highlights the remarkably high efficiency of switching the IMT off-resonance with few-cycle multi-THz pulses.

We do not observe fast recovery of an insulating character under any multi-THz field amplitude that yields measurable signals, in contrast to the behavior for resonant NIR excitation below threshold. Therefore, we now consider the microscopic model that has been introduced to capture the main features of the ultrafast IMT following NIR pumping [28,29,37].

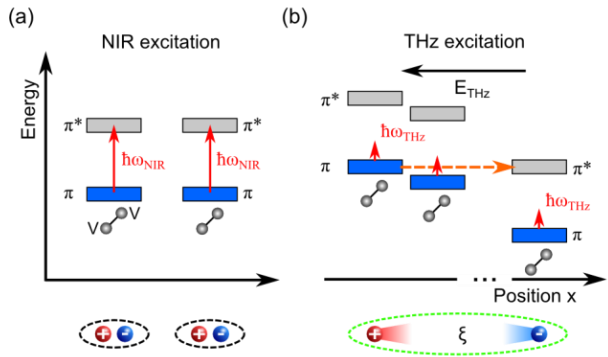

FIG. 5. Scenarios for NIR and multi-THz excitation. Red arrows denote photon energies of the excitation: (a) NIR excitation triggers a resonant interband transition. Excitonic charges are generated locally with respect to each other, resulting in neutral V-V dimers. (b) The low photon energy of the multi-THz transients keeps electrons from directly crossing the band gap. Instead, the high electric field tilts the potential enabling the generation of spatially separated doublon-holon pairs by interband tunneling (orange dashed arrow). This process leads to charged V-V dimers and a suppression of excitonic self-trapping.

To a first approximation, insulating VO$_2$ may be regarded as a set of weakly coupled V-V dimers embedded in a matrix of oxygen octahedra (Fig. 5). The valence and conduction band states are approximately described by bonding and anti-bonding superpositions of the 3$d$ orbitals of vanadium ions. Resonant NIR excitation promotes electrons into the $\pi^*$ state, leaving holes in the valence band and creating doublon-holon pairs localized around vanadium dimers, as shown schematically in Fig. 5(a). Strong doublon-holon attraction leads to exciton formation, while structural deformation of the lattice quenches the mobility of the charge carriers due to exciton self-trapping [38], a process also known in linear chain organic complexes [39]. This results in fast decay of metallic conductivity, as evidenced by the lowest graph (black) in Fig. 3(a) [28-31]. Cooperative switching into the metastable metallic phase becomes possible only above a critical density of excited dimers [28,29,37,40]. Under NIR pumping at a photon energy of $\hbar\omega_{NIR} = 1.5$ eV, it is necessary to exceed a temperature-dependent threshold fluence in the order of 5 mJ/cm$^2$ to generate metastable metallic conductivity. In stark contrast, off-resonance multi-THz pumping at a photon energy of $\hbar\omega_{NIR} = 0.1$ eV creates the long-lived metallic state even for an incident fluence of 1.6 mJ/cm$^2$ [lowest red graph in Fig. 3(a)]. The corresponding energy absorbed in the sample is far below what is necessary to heat a VO$_2$ film above $T_c$ [29], further supporting the qualitatively different character of this non-thermally induced metallization.

Since resonant absorption of multi-THz radiation is forbidden by the large energy gap, the field dependence already suggests that charge carriers are generated by tunneling as sketched in Fig. 5(b). The reason for the surprising contrast to resonant interband excitation in the near infrared is then explained by the inherent *spatial* separation of charge carriers produced via the tunneling mechanism: holons and doublons that emerge in the $\pi$ and $\pi^*$ bands, respectively, must reside on *different* V-V dimers. Under such conditions, much weaker Coulomb attraction cannot stabilize a strongly bound excitonic state. Consequently, tunneling excitation is efficient because it bypasses exciton self-trapping and quenching of conductivity in VO$_2$ that occur after resonant NIR excitation at low fluence. This conclusion is further supported by the theoretical prediction that the charge carriers induced by the Schwinger effect are energetic and almost uniformly distributed in the upper Hubbard band [11,41]. Moreover, the strong electric field of the multi-THz transients leads to additional separation of charge carriers. For weak THz fields, a potential difference comparable to the energy gap is reached on length scales that can exceed the distance between neighboring vanadium dimers. Therefore, the probability of exciton self-trapping is reduced but the number of free charge carriers decreases drastically. This is consistent with the experimental observation of long-lived conductivity excited by moderate fields.

Based on the measurement of the threshold field, we can now estimate the pair correlation length in the dielectric ground state of $VO_2$. The numerical simulations can be reproduced rather well by an analytic expression for field-dependent conductivity [10,42]

$$\sigma(E) = \sigma_\infty \exp\left(-\pi \frac{E_{th}}{E}\right), \qquad (2)$$

where $E_{th}$ is the threshold field and $\sigma_\infty$ is the maximal conductivity achieved in saturation. Indeed, Fig. 3(d) shows that Eq. (2) tracks the experimental data remarkably well with $E_{th} = 14.2$ MV/cm and $\sigma_\infty = 6084$ $\Omega^{-1}$cm$^{-1}$. This result is surprising given the complexity of correlation effects in $VO_2$ that go far beyond the Hubbard model [8-10,42]. After structural deformation, the optical gap of $VO_2$ is $\Delta = 0.6$ eV [33]. Using Eq. (1), we find a correlation length of $\xi = 2.1$ Å, which agrees well with an estimate made by Pergament [13], although a more elaborate many-body calculation is needed to substantiate this finding.

Furthermore, we can estimate a many-body analog of Keldysh's adaibaticity parameter using the following expression [11]

$$\gamma = \frac{\hbar\Omega}{\xi E}, \qquad (3)$$

where $\Omega$ is the frequency of this driving multi-THz field. For typical central frequencies of $\Omega = 22$ THz and an intermediate electrical field $E = 10$ MV/cm used in our experiments, we obtain $\gamma = 0.43$ indicating that excitation occurs in the quantum tunneling regime ($\gamma < 1$), further supporting our interpretation. It is interesting to note that for increasing frequencies and lower electric fields it would be possible to study the Keldysh crossover ($\gamma \approx 1$) between the tunneling and multiphoton excitation regimes in Mott insulators [11]. Until now, such experiments have been performed only in conventional semiconductors [43,44].

## III. SUMMARY

We have demonstrated an ultrafast insulator-to-metal transition in $VO_2$ induced by high-field multi-THz transients. The non-thermal character of the switching mechanism is confirmed by a fast onset time below 100 fs and the observation of a coherent lattice oscillation accompanying the IMT. Regardless of the excitation amplitude, the delocalized charge carriers remain mobile for hundreds of ps indicating a suppression of excitonic self-trapping due to the spatial separation of doublon and holon intrinsic to the tunneling mechanism. The density of the field-induced metallic phase is well described by theoretical models based on nonadiabatic tunneling and the Schwinger effect. The proposed microscopic mechanism of the phase transition driven by intense THz fields may also be highly relevant for efficient control of the electron delocalization in a broad class of strongly correlated transition metal oxides. Finally, we show that measuring the threshold field for tunneling breakdown gives access to the pair correlation length, a key parameter characterizing dielectric systems with electronic energy gaps due to many-body effects.

## ACKNOWLEDGEMENTS

This work has been funded by ERC Advanced Grant 290876 "UltraPhase". R.E.M. and R.F.H. acknowledge research support from the National Science Foundation (DMR-1207507). Samples were fabricated and characterized at the Vanderbilt Institute of Nanoscale Science and Engineering using facilities renovated under NSF ARI-R2 DMR-0963361.

# Tunneling Breakdown of a Strongly Correlated Insulating State in VO$_2$ Induced by Intense Multi-Terahertz Excitation

## – *Supplementary Material* –

B. Mayer,[1] C. Schmidt,[1] A. Grupp,[1] J. Bühler,[1] J. Oelmann,[1] R. E. Marvel,[2] R. F. Haglund Jr.,[2] T. Oka,[3] D. Brida,[1] A. Leitenstorfer,[1,†] and A. Pashkin[1,*]

[1] Department of Physics and Center for Applied Photonics, University of Konstanz, D-78457 Konstanz, Germany.

[2] Department of Physics and Astronomy, Vanderbilt University, Nashville, Tennessee 37235, USA.

[3] Department of Applied Physics, University of Tokyo, Bunkyo, Tokyo 113-8656, Japan.

[*]Current address Helmholtz-Zentrum Dresden-Rossendorf, Institute of Ion Beam Physics and Materials Research, 01328 Dresden, Germany.

[†]Corresponding author: aleitens@uni-konstanz.de

## I.     SAMPLE

The sample under study is a 200-nm-thick film of polycrystalline VO$_2$ grown by pulsed laser deposition on a diamond substrate synthesized by chemical vapor deposition (CVD) [S1]. A vanadium metal target is ablated in 10 mTorr of oxygen by a KrF excimer laser (repetition rate 25 Hz, nominal pulse duration 30 ns) at a wavelength of 248 nm and a fluence of 4 mJ/cm². The as-grown films are nonstoichiometric with an approximate composition close to VO$_{1.7}$. Therefore, the film is subsequently annealed at 450°C in 250 mTorr of oxygen pressure. After annealing, the VO$_2$ film phase transition was verified by temperature resolved optical spectroscopy.

.

## II. EXPERIMENTAL SETUPS

Fig. S1 depicts the setups for multi-THz pump / NIR probe [panel (a)] and multi-THz pump / multi-THz probe [panel (b)] developed for the study of the isolator-metal transition in $VO_2$. Our high-field THz source consists of a hybrid laser system combining the stability and flexibility of Er:fiber laser technology with high-power scalability of Ti:sapphire amplifiers [S2,S3]. A mode-locked Er:fiber master oscillator, working at a repetition rate of 49 MHz, is used to seed three parallel Er:fiber amplifier branches. The output of the first amplifier branch is frequency-doubled and serves as a stable seed for a high-power regenerative Ti:sapphire amplifier running at a repetition rate of 1 kHz. The laser output with energy of 5 mJ is split to pump two optical parametric amplifier (OPA) units, which generate near-infrared pulses with a tunable wavelength offset. Phase-stable multi-THz transients are generated by means of difference frequency mixing (DFG) of the OPA signal pulses in a GaSe emitter crystal. The THz radiation is focused onto the $VO_2$ sample via an off-axis parabolic mirror and subsequently characterized by electro-optic sampling (EOS) with a broadband near-infrared gate pulse of a duration of 8 fs [S4]. The ultrashort near-infrared probe and gate pulses are obtained from individually compressed supercontinua generated in highly nonlinear fibers seeded by the two remaining Er:fiber amplifier branches. This configuration results in an attosecond timing jitter [S5].For the experiments with the near-infrared probe, an 8-fs-short probe beam is collinearly overlapped with the multi-THz radiation in advance to the sample as shown in Fig. S1(a). After transmission through the sample the spectrally integrated relative transmission change is read out via a photodiode (PD). A set of 230 µm (broadband) or 1050 µm (narrowband) GaSe emitter crystals and a 140 µm GaSe detector crystal are used for the data shown in Fig. 2(a-c) in the main body of this paper. The central frequency of the pump is tuned to 22 THz with a spectral bandwidth (FWHM) of 7.7 THz and 4.5 THz, respectively. The focal diameter on the sample amounts to 62 µm. The near-infrared probe is centered at a wavelength of 1.2 µm covering a spectral bandwidth of 470 nm (FWHM) at a fluence of 10 µJ/cm$^2$ (see Fig. S2(a)). In the THz probe setup shown in Fig. S2(b) the idler signals of the OPA units are exploited for a second THz generation branch in a separated DFG stage. This scheme enables individual tuning of the center frequency as well as the pulse duration for both the multi-THz pump and multi-THz probe.

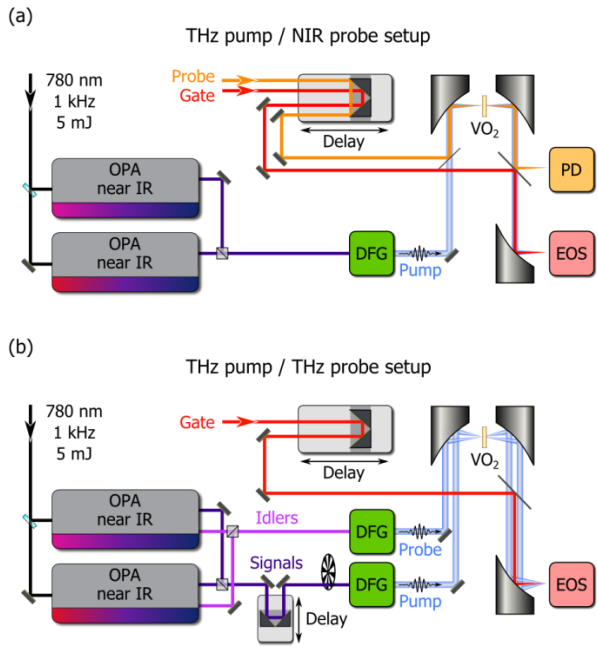

FIG. S1. (a) Scheme of the THz pump / NIR probe setup. The signals of two OPAs at variably frequency are superimposed for DFG in a GaSe crystal. The multi-THz radiation (light blue) is focused on the $VO_2$ sample via an off-axis parabolic mirror and then electro-optically sampled by a near-infrared gate pulse (red). A broadband probe pulse (orange) is collinearly overlapped with the THz radiation before the sample. The transmission change is read out via a photodiode (PD). (b) Scheme of the THz pump / THz probe setup: In addition to the system depicted in (a) the idler signals of the OPA units are used for a second THz generation branch serving as the probe.

The pump is modulated by a mechanical chopper with a frequency of 500 Hz. The two phase-locked multi-THz pulses are aligned non-collinearly and focused on the $VO_2$ sample. The transmitted signals of pump and probe are simultaneously detected by electro-optic sampling (EOS) with a sub-cycle precision. Here the pump is generated by a 660-µm-thick GaSe and the probe by a 3-mm-thick GaSe emitter crystal. Both components are sampled in a 140 µm GaSe detector. The pump is centered at 28 THz with a spectral bandwidth of 5.9 THz (FWHM), the probe is centered at 25 THz with a FWHM of 4.5 THz [see Fig. S2(b)]. Due to the non-collinear alignment, the focal spot sizes are slightly larger than that of the THz pump / NIR probe conditions. The fluence of the probe beam is 0.8 mJ/cm$^2$ corresponding to an internal field of 2 MV/cm. For the data shown in the main body of the contribution, the delay time between pump and probe is set to be 1 ps.

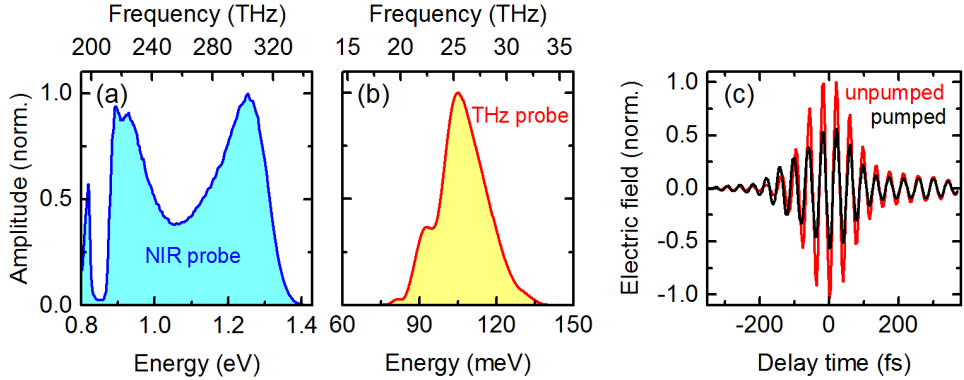

FIG. S2: Amplitude spectrum of the (a) NIR and (b) THz probe beams. (c) Electro-optically sampled multi-THz probe transient transmitted through the 200 nm $VO_2$ sample in absence of the pump (red). After pumping the sample by a multi-THz transient centered at 28 THz (not shown) with a peak electric field of 15.9 MV/cm, the transmission of the probe beam is drastically suppressed (black). Delay time between pump and probe is set to 1 ps.

Fig. S2(c) shows the multi-THz probe transient with a duration of 153 fs (FWHM) transmitted through the unperturbed $VO_2$ sample at room temperature, as analyzed via electro-optic detection [red graph, corresponding spectrum is shown in Fig. 2(b)]. After pumping the sample with a multi-THz transient centered at 28 THz and a peak electric field of 15.9 MV/cm (not shown) the transmission of the probe field is suppressed by more than 50% at a pump-probe delay of 1 ps (black curve). It is clearly seen that the pump-induced changes occur in phase with the multi-THz probe. The relative transmission change depicted in Fig. 3(d) in the main body of the paper is extracted from the change of spectral weight of the probe pulse. Therefore, the total spectral intensity for the unpumped

$$A_{\text{unpumped}} = \int_0^\infty \left|\text{FT}[E(t)_{\text{unpumped}}]\right|^2 d\omega,$$

and pumped

$$A_{\text{pumped}} = \int_0^\infty \left|\text{FT}[E(t)_{\text{pumped}}]\right|^2 d\omega,$$

cases are determined and the relative transmission change is calculated by the following expression

$$\frac{\Delta T}{T} = \frac{A_{\text{pumped}} - A_{\text{unpumped}}}{A_{\text{unpumped}}}$$

## III. METASTABLE METALLIC STATE INDUCED BY MULTI-THz TRANSIENTS

Fig. S3 shows a THz pump / NIR probe signal measured over a large range of pump-probe delay times reaching 80 ps. The excitation pulse with a peak field of 7.0 MV/cm (incident fluence of 14.9 mJ/cm$^2$) centered at 30.5 THz induces a relatively weak decrease in the NIR transmission. Nevertheless, this pump-induced change does not even show significant relaxation on a time scale up to 80 ps. This result provides solid evidence for the metastable character of the metallic conductivity in VO$_2$ induced by multi-THz pulses. We have studied the full relaxation to the insulating state directly on a fast oscilloscope and found that it occurs on timescales of a few nanoseconds.

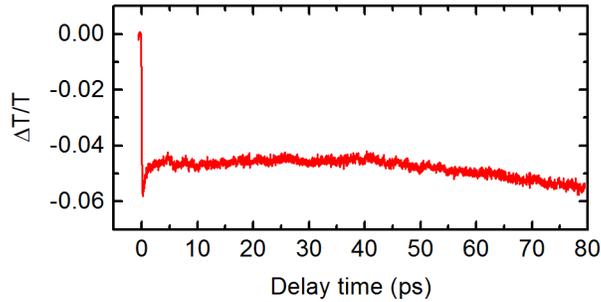

FIG. S3: Relative transmission change $\Delta T/T$ of the 200-nm-thick VO$_2$ film induced by a multi-THz transient with peak electric field of 7.0 MV/cm.

**REFERENCES**

[S1] J. Y. Suh, R. Lopez, L. C. Feldman, and R. F. Haglund, Jr., Appl. Phys. **96**, 1209 (2004).
[S2] A. Sell, A. Leitenstorfer, and R. Huber, Opt. Lett. **33**, 2767 (2008).
[S3] A. Pashkin, F. Junginger, B. Mayer, C. Schmidt, O. Schubert, D. Brida, R. Huber, and A. Leitenstorfer, IEEE J. Sel. Top. Quantum Electron. **19**, 8401608 (2013).
[S4] A. Sell, G. Krauss, R. Scheu, R. Huber, and A. Leitenstorfer, Opt. Express **17**, 1070 (2009).
[S5] F. Adler, A. Sell, F. Sotier, R. Huber, and A. Leitenstorfer, Opt. Lett. **32**, 3504 (2007).